\documentclass[aps,prb,a4paper,twocolumn,noshowpacs,superscriptaddress,floatfix]{revtex4}
\usepackage{graphicx}
\usepackage{amsmath}
\usepackage{amssymb}

\begin{document}
\title{Influence of Roughness and Disorder on Tunneling Magnetoresistance}
\author{P. X. Xu}
\affiliation{Beijing National Laboratory for Condensed Matter Physics,
Institute of Physics, Chinese Academy of Sciences, Beijing 100080,
China}
\author{V. M. Karpan}
\affiliation{Faculty of Science and Technology and MESA$^+$ Institute
for Nanotechnology, University of Twente, P.O. Box 217, 7500 AE
Enschede, The Netherlands}
\author{K. Xia}
\affiliation{Beijing National Laboratory for Condensed Matter Physics,
Institute of Physics, Chinese Academy of Sciences, Beijing 100080,
China}
\author{M. Zwierzycki}
\affiliation{Faculty of Science and Technology and MESA$^+$
Institute for Nanotechnology, University of Twente, P.O. Box 217,
7500 AE Enschede, The Netherlands}
\affiliation{Max-Planck-Institut f\"{u}r Festk\"{o}rperforschung,
Heisenbergstr. 1, D-70569 Stuttgart, Germany.}
\author{I. Marushchenko}
\affiliation{Faculty of Science and Technology and MESA$^+$ Institute
for Nanotechnology, University of Twente, P.O. Box 217, 7500 AE
Enschede, The Netherlands}
\author{P. J. Kelly }
\affiliation{Faculty of Science and Technology and MESA$^+$
Institute for Nanotechnology, University of Twente, P.O. Box 217,
7500 AE Enschede, The Netherlands}
\date{\today}

\begin{abstract}
A systematic, quantitative study of the effect of interface
roughness and disorder on the magnetoresistance of
FeCo$|$vacuum$|$FeCo magnetic tunnel junctions is presented based
upon parameter-free electronic structure calculations. Surface
roughness is found to have a very strong effect on the
spin-polarized transport while that of disorder in the leads
(leads consisting of a substitutional alloy) is weaker but still
sufficient to suppress the huge tunneling magneto-resistance (TMR)
predicted for ideal systems.
\end{abstract}


\maketitle

Tunneling magnetoresistance (TMR) refers to the dependence of the
resistance of a FM$_1|$I$|$FM$_2$
(ferromagnet$|$insulator$|$ferromagnet) magnetic tunnel junction (MTJ)
on the relative orientation of the magnetization directions of the
ferromagnetic electrodes when these are changed from being antiparallel
(AP) to parallel (P): TMR $= (R_{AP}-R_{P})/R_P \equiv
(G_{P}-G_{AP})/G_{AP}$. Since the discovery of large values of TMR in
MTJs based upon ultrathin layers of amorphous Al$_2$O$_3$ as insulator,
\cite{Moodera:prl95} a considerable effort has been devoted to
exploiting the effect in sensors and as the basis for non-volatile
memory elements. Understanding TMR has been complicated by the
difficulty of experimentally characterizing FM$|$I interfaces. The
chemical composition of the interface has been
shown\cite{DeTeresa:sc99} to have a strong influence on the magnitude
and polarization of the TMR and knowledge of the interface structure is
a necessary preliminary to analyzing MTJs theoretically. In the absence
of detailed structural models of the junctions and the
materials-specific electronic structures which could be calculated with
such models, the effect was interpreted in terms of electrode
conduction-electron spin polarizations $P_{i}$, using a model suggested
by Julliere \cite{Julliere:pla75} in which the TMR $=2P_1 P_2/(1- P_1
P_2)$. A great deal of discussion has focussed on the factors
contributing to the quantity \cite{note1} $P$ but the use of amorphous
oxide as barrier material made impossible a detailed theoretical study
with which to confront experiment.\cite{Tsymbal:jpcm03,Zhang:jpcm03}

The situation changed quite drastically with the recent observation of
large values of TMR at room temperature in FeCo$|$MgO$|$FeCo MTJs in
which the MgO tunnel barrier was mono-\cite{Yuasa:natm04,Yuasa:apl05}
or poly-crystalline.\cite{Parkin:natm04} This work was motivated in
part by the prediction\cite{Butler:prb01,Mathon:prb01} by
materials-specific transport calculations of huge TMR values for ideal
Fe$|$MgO$|$Fe structures. This new development lends fresh urgency to
the need to understand the factors governing the sign and magnitude of
TMR because the largest observed value of 353\% at low
temperature,\cite{Yuasa:apl05} is still well below the ab-initio
predicted values of order 10,000\% for the relevant thicknesses of
MgO.\cite{Butler:prb01} Some effort has been devoted to explaining the
discrepancy in terms of interface relaxation\cite{Wortmann:jpcm04} or
the formation of a layer of FeO at the
interface\cite{Zhang:prb03,Tusche:prl05} but the role of interface
disorder has only been speculated upon.\cite{Belashchenko:prb05b}

\underline{Method}. In this paper, we use first principles electronic
structure calculations to study the effect of roughness and alloy
disorder on TMR in MTJs with a vacuum barrier and Fe$_{1-x}$Co$_x$
alloy electrodes. A vacuum barrier was chosen for its simplicity and
because there are many studies of spin-dependent vacuum tunneling in
its own right.%
\cite{Stroscio:prl95,Alvarado:prl95,Okuno:prl02,Ding:prl03,Bischoff:prb03}
We consider the effect of diffusive scattering in the linear-response
regime in a two-step procedure. In the first step, the electronic
structure of the Fe$_{1-x}$Co$_x|$vacuum$|$Fe$_{1-x}$Co$_x$ MTJ is
determined using the local-density approximation \cite{vonBarth:jpc72}
of density functional theory. The self-consistent calculations are
performed with the tight-binding linear muffin-tin orbital (TB-LMTO)%
\cite{Andersen:prb86} surface Green's function
method\cite{Turek:97} and disordered systems are treated using the
layer CPA (coherent potential approximation).\cite{Soven:pr67} The
atomic sphere (AS) potentials serve as input to the second step in
which the transmission matrix entering Landauer's transport formalism%
\cite{Datta:95} is calculated using a TB-MTO implementation%
\cite{Xia:prb01,Xia:prb06} of a wave-function matching scheme due
to Ando.\cite{Ando:prb91} Disorder is modelled in large lateral
supercells by distributing the self-consistently calculated CPA-AS
potentials at random, layer-for-layer in the appropriate
concentrations for as many configurations as are required. In most
of the calculations to be presented, supercells containing $10
\times 10$ atoms per monatomic layer were used.

We consider transport in the (001) growth direction keeping atoms at
the surfaces unrelaxed in their bulk bcc positions. For Fe leads the
experimental lattice constant $a_{\mathrm{Fe}}=2.866$\AA~ is used. For
Co leads $a_{\mathrm{Co}}=2.817$\AA~ is chosen so that the bcc lattice
has the same volume as hcp Co. The alloy is considered to obey Vegard's
law whereby
\begin{equation}
a_{\mathrm{Fe_{1-x}Co_x}}=(1-x) a_{\mathrm{Fe}}+ x a_{\mathrm{Co}}
\label{eq1}
\end{equation}
The vacuum region is modelled in the atomic spheres approximation (ASA)
by filling the space between the leads with `empty' spheres of the same
size \cite{fn:WS_radii} and on the same bcc lattice as in the leads.
Consequently we measure the thickness of the vacuum barrier in
monolayers (MLs) of empty atomic spheres.

\begin{figure}[t]
\includegraphics[scale=0.67]{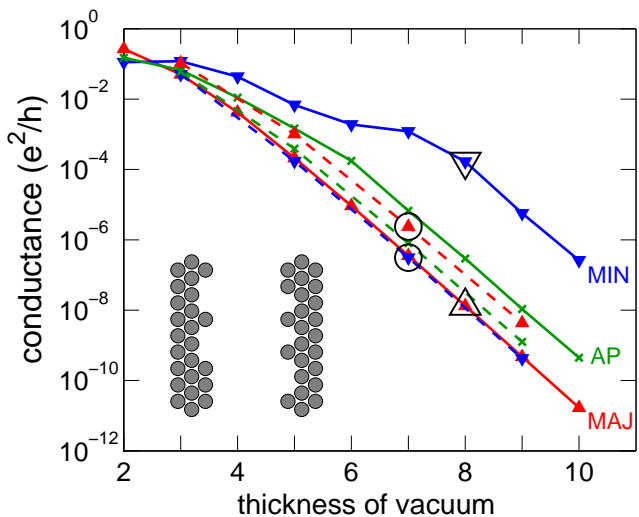}
\caption{(Color online)
Conductances
$G_P^{min}$ ($\blacktriangledown$),
$G_P^{maj}$ ($\blacktriangle$), and
$G_{AP}^{\sigma}$ (${\mathbf \times}$) of an Fe$|$vacuum$|$Fe MTJ as
a function of the barrier thickness (measured in units of layers of a
bcc lattice). The solid lines are for ideal junctions. The dashed lines
are configuration-averaged conductances for rough junctions prepared by
removing, at random, half of the Fe atoms from one surface and
depositing them, at random, on the other surface. The results are
normalized to the $1 \times 1$ surface unit cell used for the ideal
case. The large symbols refer to data points which also appear in the
next figure.}
\label{Fig1}
\end{figure}

\underline{Ideal Fe$|$vac$|$Fe}. The conductances $G_P^{\sigma}$ and
$G_{AP}^{\sigma}$ for an ideal, ordered Fe$|$vacuum$|$Fe MTJ are shown
in Fig.~\ref{Fig1} (solid lines) as a function of the width of the
barrier for the minority and majority spin channels, $\sigma= min,
maj$. In all three cases an exponential dependence on the barrier width
is reached asymptotically. For $G_P^{maj}$, the asymptotic dependence
sets in after about five MLs of vacuum; the initial subexponential
behaviour is related to the deviation of the barrier from a simple step
form. The behaviour of $G_P^{min}$ is much more complex, becoming
exponential only when the vacuum is some 8 MLs or more thick; the AP
case is intermediate. The TMR increases\cite{fn:pol} as a function of
the width of the vacuum barrier reaching huge asymptotic values of
order 20,000\%, similar in size to those
reported\cite{Butler:prb01,Mathon:prb01} for Fe$|$MgO$|$Fe. The
non-trivial dependence of $G_P^{min}$ on barrier width can be
understood in terms of a surface state\cite{Stroscio:prl95} at
$\bar{\Gamma}$ (${\bf k}_{\parallel}=0$) which is very close to the
Fermi level in the minority spin channel of iron but well below it for
majority spin electrons. In the P configuration one such state exists
on each surface and at values of ${\bf k}_{\parallel} \neq 0$ these
states become surface resonances which form bonding-antibonding pairs
with transmission probabilities close to unity.\cite{Wunnicke:prb02a}
As the vacuum barrier is widened, the coupling between the surface
resonances weakens until the bonding-antibonding splitting becomes
smaller than the resonance width at which point the maximum
transmission becomes smaller than unity and normal exponential
dependence of the conductance on the barrier width sets in; the
asymptotic slope in Fig.~\ref{Fig1} is consistent with our calculated
Fe workfunction of 4.7 eV.

\underline{Roughness: 50\% coverage}. An ideal tunnel junction such as
that just considered is impossible to realize in practice; there will
always be some finite amount of disorder whether it be surface
roughness, islands, dislocations etc. A more realistic model is
obtained by considering an Fe$|$vacuum$|$Fe system with incomplete
(rough) surface layers, modelled by occupying, at random, a fraction of
the lattice sites of the topmost layer with Fe atoms (see the inset in
Fig.~\ref{Fig1}) and using the layer CPA to determine the corresponding
AS potentials.

Roughness resulting from depositing some Fe atoms on an ideal surface
has two effects. The first is to destroy the point group symmetry which
led to the existence of a symmetry gap at $\bar{\Gamma}$ for states
with $\Delta_1$ symmetry, and thus the surface state with that symmetry.
We expect this to reduce the large minority spin conductance found for
the ideal vacuum barrier. The second effect of roughness is to reduce
the width of the vacuum barrier and thus to enhance the conductance
which depends exponentially on the barrier width.

These competing effects are disentangled by starting with an ideal
vacuum barrier and moving half of the Fe atoms from one surface and
depositing them on the other so that the barrier width is, on average,
unchanged. The results of these supercell calculations, averaged over
20 configurations and normalized to the ideal MTJ ($1 \times 1$ surface
unit cell) results are included in Fig.~\ref{Fig1}. The contribution of
the interface states to $G_P^{min}$ is completely quenched by the
roughness and the conductance reduced by four orders of magnitude.
$G_P^{maj}$ is enhanced by about an order of magnitude because the
increased conductance from those parts of the rough surfaces which are
closer than average more than compensates the decreased conductance
from part of the rough surfaces which are further away than average.
All conductances now exhibit the same qualitative behavior: exponential
decay. Not only is the absolute value of the TMR greatly reduced by
roughness, the sign of the polarization is reversed.

\begin{figure}[!]
\includegraphics[scale=0.67]{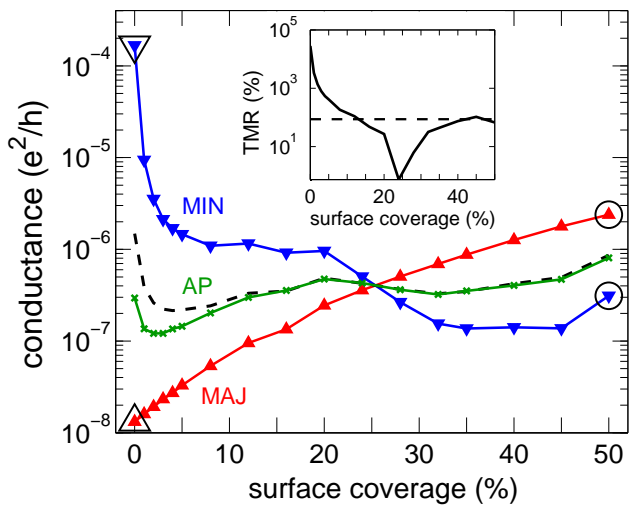}
\caption{(Color online) Configuration-averaged conductances $G_P^{min}$
($\blacktriangledown$), $G_P^{maj}$ ($\blacktriangle$), and
$G_{AP}^{\sigma}$ (${\mathbf \times}$) of an Fe$|$vacuum$|$Fe MTJ with
8 ML thick vacuum barrier as a function of the surface coverage,
normalized to a $1 \times 1$ surface unit cell. The dashed line denotes
$G_{AP}^{\sigma}$ predicted from Eq.~\ref{eq2}. Large symbols refer to the 
similarly marked data points in the previous figure.
Inset: TMR as a function of the surface coverage. The dashed line is the value
predicted using Julliere's expression and a calculated DOS polarization
of 55\%. } \label{Fig2}
\end{figure}

This result immediately prompts us to ask how the TMR depends on the
amount of roughness and, in particular, what coverage is needed to
suppress the contribution of the resonances to $G_P^{min}$. This issue
is addressed in Fig.~\ref{Fig2} where the conductance of a MTJ is shown
as a function of surface coverage. Zero coverage corresponds to an
ideal MTJ with 8 MLs vacuum; 100\% coverage (not shown) to 6 MLs. We
see that 5\% coverage is sufficient to reduce $G_P^{min}$, and the TMR,
by two orders of magnitude. Unless the surface roughness is less than a
few percent, the TMR lies between about 10 and 1000\% (Fig.~\ref{Fig2},
inset), comparable to values observed in experiment. As $G_P^{min}$ is
reduced by roughness, $G_P^{maj}$ increases monotonically with
increasing coverage as the average barrier width decreases; the two
cross at a surface coverage of about 25$\%$.

The single spin AP conductance, $G_{AP}^{\sigma}$, is described
qualitatively very well for all coverages by the
relation\cite{Belashchenko:prb04}
\begin{equation}
 G_{AP}^{\sigma}=\sqrt{G^{maj}_P G^{min}_P}
\label{eq2}
\end{equation}
(dashed line in Fig.~\ref{Fig2}) and quantitatively for coverages
greater than a few percent i.e., as soon as the surface resonance
contribution is killed by roughness.\cite{fn:MU} When Eq.~\eqref{eq2}
holds, $G_P \equiv G^{maj}_P + G^{min}_P$ is greater than $G_{AP}
\equiv 2G^{\sigma}_{AP} $ and the TMR is always positive although the
polarization in the P configuration changes sign, reaching its minimum
value at 25$\%$ coverage where $G^{maj}_P$ and $G^{min}_P$ crossover.

As long as there is enough roughness to quench the surface resonance,
there is order-of-magnitude agreement between our TMR calculated as a
function of surface roughness and the value obtained using Julliere's
expression for the TMR with the calculated, bulk density-of-states
polarization for Fe, $P = 55\%$, shown as a dashed line in the inset to
Fig.~\ref{Fig2}.

\underline{Disorder in the leads}. It is interesting to study an
intermediate type of disorder where there is no roughness but the
electrodes are made of a substitutional Fe$_{1-x}$Co$_x$ magnetic
alloy. As the alloy concentration $x$ is increased, the disorder
increases but the underlying electronic structure is also changing as
the Fermi energy rises. To distinguish these two effects, we first
carry out calculations for a VCA$|$vac$|$VCA MTJ in which the
Fe$_{1-x}$Co$_x$ electrodes are treated within the virtual crystal
approximation (VCA). This allows us to probe the effect of changing the
Fermi energy without including disorder. It turns out to be also very
convenient to embed Fe$_{1-x}$Co$_x|$vac$|$Fe$_{1-x}$Co$_x$ between VCA
leads (schematically denoted) as VCA$|$CPA$|$vac$|$CPA$|$VCA for
performing the self-consistent potential calculations and as
VCA$|$SC$|$vac$|$SC$|$VCA for the conductance calculation where, as
before, CPA potentials are used as input to the supercell (SC)
transport calculations. Details of the thicknesses of the various
regions needed to achieve converged results as well as the number of
k-points, supercell sizes and other technical details will be given in
a forthcoming publication.

\begin{figure}[!]
\includegraphics[scale=0.67]{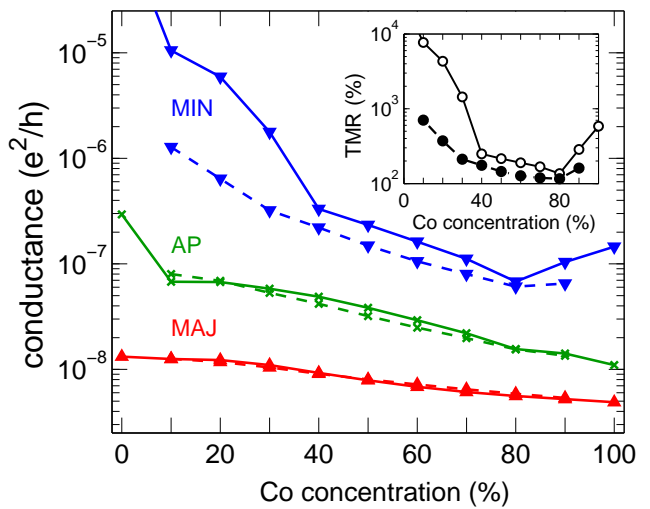}
\caption{(Color online) $G_P^{min}$ ($\blacktriangledown$), $G_P^{maj}$
($\blacktriangle$), and $G_{AP}^{\sigma}$ (${\mathbf \times}$) for an
Fe$_{1-x}$Co$_x|$vacuum$|$Fe$_{1-x}$Co$_x$ MTJ with 8 ML thick vacuum
barrier as a function of $x$, the concentration of Co atoms, calculated
in the virtual crystal (VCA: solid lines) and CPA/supercell (SC: dashed
lines) approximations. Conductances are configuration averaged and
normalized to a $1 \times 1$ surface unit cell. Inset: TMR in VCA
($\circ$) and CPA/SC ($\bullet$) approximations.}
\label{Fig3}
\end{figure}

The results of these calculations are shown in Fig.~\ref{Fig3}.
$G_P^{maj}$ and $G_{AP}^{\sigma}$ are unchanged by disorder within the
accuracy of the calculation. The largest change can once again be seen
in $G_P^{min}$. In the case of ideal VCA electrodes, there are
localized surface states in the minority channel as one would expect.
With increasing Co concentration $x$, the Fermi level rises with
respect to these surface states, resulting in a decreasing contribution
to the conductance from the surface resonances. This is clearly
demonstrated by the behavior of $G_P^{min}$ in Fig.~\ref{Fig3}. Lead
disorder quenches the contribution of the resonant states by destroying
the point group symmetry, eliminating the symmetry gap and broadening
the resonances. At a concentration of about 40\%, the surface
resonances are no longer dominant and the effect of disorder is quite
small; for $x \geq 0.4$ the trend as a function of $x$ is described
quite well by the VCA calculation. Eq.~\eqref{eq2} is again found to
describe the AP conductance very well. We see thus that lead disorder
has the same effect as roughness in quenching the huge values of TMR
(inset to Fig.~\ref{Fig3}) found for the disorder-free VCA reference
system, albeit less effectively.

\underline{Summary}. Using first-principles calculations we find that
both roughness and alloy disorder quench the TMR for ideal magnetic
tunnel junctions to values comparable to the highest found
experimentally for monocrystalline barrier materials or vacuum
tunneling. We therefore propose that these experiments are still in the
roughness/disorder limited regime.

This work is part of the research program of the ``Stichting voor
Fundamenteel Onderzoek der Materie'' (FOM) and the use of supercomputer
facilities was sponsored by the ``Stichting Nationale Computer
Faciliteiten'' (NCF), both financially supported by the ``Nederlandse
Organisatie voor Wetenschappelijk Onderzoek'' (NWO). VMK is supported
by ``NanoNed'', a Nanotechnology programme of the Dutch Ministry of
Economic Affairs. KX acknowledges support from NSF of China, Grant
No.~90303014.

We are grateful to Anton Starikov for permission to use his version of
the TB-MTO code based upon sparse matrix techniques with which we
performed some of the calculations.


\begin{thebibliography}{35}
\bibitem[{not()}]{note1}
\bibinfo{note}{$P_i$ was not defined strictly in
  Ref.~\onlinecite{Julliere:pla75} and it has frequently been interpreted in
  terms of the density-of-states polarization
  $P=[D_{\uparrow}(\varepsilon_f)-D_{\downarrow}(\varepsilon_f)]/
  [D_{\uparrow}(\varepsilon_f)+D_{\downarrow}(\varepsilon_f)]$ where
  $D_{\sigma}(\varepsilon)$ is a bulk density of states. It has been pointed
  out \cite{Mazin:prl99} that other measures of the spin polarization may be
  more appropriate when studying transport.}
\bibitem[{fn:({\natexlab{c}})}]{fn:MU}
\bibinfo{note}{For ideal MTJs, the resonant tunneling contribution will be
  quenched for sufficiently thick barriers so Eq.~\eqref{eq2} can also be
  expected to hold in that situation.}
\end{thebibliography}

\end{document}